\documentclass[twocolumn,showpacs,preprintnumbers,amsmath,amssymb,prl]{revtex4}

\usepackage{graphicx}
\usepackage{dcolumn}
\usepackage{bm}

\usepackage{mathptmx, courier, pifont}
\usepackage[scaled=0.92]{helvet}
\usepackage[T1]{fontenc}
\usepackage{textcomp}

\newcommand{\quarterthin}{\kern 0.0417em}

\begin{document}


\title{Isospin symmetry breaking at high spins in the mirror pair
$^{67}$Se and $^{67}$As}

\author{K.~Kaneko$^{1}$, S.~Tazaki$^{2}$, T.~Mizusaki$^{3}$,
 Y.~Sun$^{4,5}$, M.~Hasegawa$^{4,5}$, G.~de~Angelis$^{6}$ }

\affiliation{
$^{1}$Department of Physics, Kyushu Sangyo University, Fukuoka
813-8503, Japan \\
$^{2}$Department of Applied Physics, Fukuoka University, Fukuoka
814-0180, Japan \\
$^{3}$Institute of Natural Sciences, Senshu University, Tokyo
101-8425, Japan \\
$^{4}$Department of Physics, Shanghai Jiao Tong
University, Shanghai 200240, People's Republic of China \\
$^{5}$Institute of Modern Physics, Chinese Academy of Sciences,
Lanzhou 730000, People's Republic of China \\
$^{6}$Laboratori Nazionali di Legnaro dell'INFN, Legnaro (Padova),
I-35020, Italy
}

\date{\today}

\begin{abstract}

Recent experimental data have revealed large mirror energy differences
(MED) between high-spin states in the mirror nuclei $^{67}$Se and $^{67}$As,
the heaviest pair where MED have been determined so far. The MED are generally
attributed to the isospin symmetry breaking caused by the Coulomb force
and by the isospin nonconserving part of the nucleon-nucleon residual
interaction. The different contributions of the various terms have
been extensively studied in the $fp$ shell. By employing large-scale shell
model calculations, we show that the inclusion of the $g_{9/2}$ orbit causes
interference between the electromagnetic spin-orbit and the Coulomb
monopole radial terms at high spin. The large MED are attributed to the
aligned proton pair excitations from the $p_{3/2}$ and $f_{5/2}$ orbits
to the $g_{9/2}$ orbit. The relation of the MED to deformation is discussed.

\end{abstract}


\maketitle


One of the current topics in nuclear structure physics is the
isospin symmetry breaking due to the Coulomb force and the strong
nucleon-nucleon (NN) interaction.  Assuming isospin symmetry,
mirror pair nuclei, i.e. a pair of nuclei with exchanged proton and
neutron numbers, have identical level schemes.  However, the Coulomb
effects and the isospin nonconserving NN interaction break this
symmetry, leading to observable differences between energy levels of
analogue states.  The so-called mirror energy differences (MED) are
defined by
\begin{eqnarray}
 {\rm MED}_{J} & = & E_{x}(J,T,T_{z}=-T) - E_{x}(J,T,T_{z}=T),
          \label{eq:0}
\end{eqnarray}
where $E_{x}(J,T,T_{z})$ are the excitation energies of analogue
states with spin $J$ and isospin $T,T_{z}$.  The MED are thus
regarded as a measure of isospin symmetry breaking in an effective
interaction which includes the Coulomb force.  The MED have been extensively
studied for mirror pair nuclei in the upper $sd$ and the lower
$fp$ shell regions (see Ref. \cite{Bentley07} for review). In both cases, a
remarkable agreement between experimental data and shell model
calculations has been achieved, allowing a clear identification of the
origin of the MED based on the isospin nonconserving Coulomb and
strong NN forces
\cite{Lenzi06,Jenkins05,Nara07,Bentley00,Duflo02,Lenzi01,Zuker01,
Garrett07,Garrett01,Gadea06,Vedora07,Ekman04,Bentley06,Brown09,Rudolph08,
Andersson05,Warner06}.

For mirror nuclei in the upper part of the $fp$ shell the situation is
different. The experimental information on MED is relatively scarce
and only recent large-scale shell model calculations including the
$g_{9/2}$ orbit have become available \cite{Hasegawa05,Honma09}.
Moreover due to the deformation driving effect of the $g_{9/2}$ orbit,
variations in the MED are expected to be strongly related to the change
in the nuclear deformation. Recently new data on the $A$=67 mirror
nuclei $^{67}$Se and $^{67}$As have become available \cite{Orlandi09}.
Investigations \cite{Stefanova03} for $^{66}$Ge suggested that
the spin alignment of the $g_{9/2}$ neutrons occurs at
$J^{\pi}=8^{+}$.  As the positive-parity band built on the $9/2^{+}$
state in $^{67}$As can be interpreted as a $g_{9/2}$ proton weakly
coupled to the $^{66}$Ge core, the neutron spin alignment is
expected to occur at spin $25/2^{+}(=8^++9/2^+)$ in $^{67}$As
\cite{Jenkins01,Stefanova03}.  On the other hand, the proton spin
alignment takes place at the same spin in its mirror partner
$^{67}$Se.  As the response to the Coulomb field is different
for the corresponding high-spin states in such mirror nuclei, one expects
the Coulomb based MED contribution in $^{67}$Se and $^{67}$As to
give large negative value suddenly at $25/2^{+}$ where the proton/neutron spin alignment
occurs. In the lower $fp$-shell region, due to the active
role played by the $f_{7/2}$ shell, in the MED the isospin nonconserving NN interaction
has been suggested to be at least as important as
the Coulomb part \cite{Zuker02}. In the upper $fp$-shell region the situation is
different and one does not expect a major contribution because the $f_{7/2}$
shell is almost not active.

For the $A=67$ mirror pair nuclei, excited states have been known for
$^{67}$As \cite{Grzywacz01,Jenkins01}, and have been recently determined for the
mirror partner $^{67}$Se \cite{Angelis07}.  This is the heaviest
mirror pair where the excited energy levels have been identified with detailed experimental
information.  In both cases, the low-lying $9/2^{+}$ state has been found to be isomeric,
allowing the determination of the degree of isospin symmetry breaking through the measurement
of the mirror $9/2^{+}\rightarrow 7/2^{-}$ $E1$ strengths \cite{Orlandi09}.
In our previous paper \cite{Hasegawa05}, the structure of
this isomeric state has been investigated using large-scale shell
model calculations.  The isomerism of the $9/2^{+}$ state was
understood as due to proton and neutron configuration mixing
based on the $g_{9/2}$ intruder orbit as well as on the $fp$-shell
structures.

In this Rapid Communication, we investigate the MED in the mirror pair
$^{67}$Se and $^{67}$As discussing the origin of isospin symmetry
breaking in the upper $fp$-shell region.
Theoretical calculations are performed using the spherical shell
model in the $pf_{5/2}g_{9/2}$ model space.  We employ the recently
proposed JUN45 interaction \cite{Honma09}, a
realistic effective interaction based on the Bonn-C potential and
ajusted to the experimental data of nuclei in the $A=63\sim 96$
mass region.
To describe the MED, the first attempt was carried out by adding the
Coulomb term to the KB3 interaction matrix elements
\cite{Caurier94}. However, those calculations did not succeed to
describe the experimental MED for the mirror pairs of mass $A=47$
and $A=49$. A better agreement with the data has been obtained using
the formalism introduced by Zuker {\it et al.} \cite{Zuker02}. In this
description the Coulomb matrix elements in the valence space represent
only the multipole part of the Coulomb interaction whereas the contribution
of the other nucleons is described by the Coulomb monopole effect. The
Coulomb interaction is therefore separated into a monopole term $V_{Cm}$
and a multipole term $V_{CM}$. While $V_{Cm}$ accounts for single-particle
and bulk effects, $V_{CM}$ contains all the rest. The monopole term $V_{Cm}$
is further divided into the single particle correction $\varepsilon_{ll}$,
the radial term $V_{Cr}$ and the spin orbit term $\varepsilon_{ls}$.
The contribution of $\varepsilon_{ll}$ to the monopole term is given by \cite{Duflo02}
\begin{eqnarray}
 \varepsilon_{ll} & = & \frac{-4.5{Z_{cs}^{13/12}}[2l(l+1)-p(p+3)]}{A^{1/3}(p+3/2)},
          \label{eq:1}
\end{eqnarray}
where $Z_{cs}$ is the proton number  corresponding to a closed shell,
$p$ the principal quantum number, and $l$ the orbital momentum. Due
to such single particle correction, in $^{67}$Se the proton $g_{9/2}$
and $f_{5/2}$ orbits are lowered roughly by 95 keV and 58 keV, respectively,
while the energy of the $p_{3/2}$ orbit is raised by about 135 keV.
The relative energy gap between the proton $g_{9/2}$ and $f_{5/2}$
orbits is reduced of only 37 keV, and therefore there is basically no effect
on single-particle levels due to the $\varepsilon_{ll}$ term.

The radial term $V_{Cr}$ reflects the change in radii along the
rotational band, and in the $fp$ shell is proportional to the change
in occupancy of the $p_{3/2}$ orbit as a function of spin $J$.  It
can be expressed as $\Delta_{MED}(V_{Cr})=a_{m}(\langle
m_{p3/2}\rangle_{9/2} /2-\langle m_{p3/2}\rangle_{J} /2)$, where
$\langle m_{p3/2}\rangle_{J}$ with $m_{p3/2}=z_{p3/2}+n_{p3/2}$ is
the expectation value of the proton and neutron number in the
$p_{3/2}$ orbit at spin $J$ and $a_{m}$ is the strength parameter
fitted to the experimental data. When the occupation of the
$p_{3/2}$ protons decreases, valence protons in orbits with smaller
radii are nearer to the charged core, which results in a gain of
Coulomb energy \cite{Bentley07}. In the $pf_{5/2}g_{9/2}$ shell, the
$p_{3/2}$ orbit has larger radius than the $f_{5/2}$ and $g_{9/2}$
orbits and therefore the Coulomb repulsion increases as the number
of protons increases. Here the role of the $p_{1/2}$ orbit is less
important simply because the $p_{1/2}$ occupancy is small, and
furthermore it does not change very much as a function of the
angular momentum $J$. 

The single-particle shift $\varepsilon_{ls}$ takes into account
the relativistic spin-orbit interaction
\cite{Nolen69}.  This interaction comes from the Larmor precession
of the nucleons in the electric field due to their magnetic moments,
which, as well known, affects the single-particle energy spectrum.
$\varepsilon_{ls}$ can be written as \cite{Nolen69}
\begin{eqnarray}
 \varepsilon_{ls} & = & (g_{s}-g_{l})\frac{1}{2m_{N}^{2}c^{2}}\left( \frac{1}{r}\frac{dV_{c}}{dr}
  \right)\langle \hat{l}\cdot\hat{s}\rangle,
          \label{eq:2}
\end{eqnarray}
where $m_{N}$ is the nucleon mass, and the free values of the
gyromagnetic factors, $g_{s}^{\pi}$=5.586, $g_{l}^{\pi}$=1 for
protons and $g_{s}^{\nu}$=-3.828, $g_{l}^{\nu}$=0 for neutrons, are
used. In the present work, by assuming a uniformly charged
sphere, $\varepsilon_{ls}$ is calculated using the harmonic
oscillator single-particle wave function. Depending on proton or
neutron orbit, the shift can have opposite signs.  It depends also
on the spin-orbit coupling, as for instance $\langle
\hat{l}\cdot\hat{s}\rangle =l/2$ when $j=l+s$ and $\langle
\hat{l}\cdot\hat{s}\rangle =-(l+1)/2$ when $j=l-s$.  As this term
influences differently on neutrons and protons, its effect becomes
very important for some particular states.  In $^{67}$Se, the proton
$g_{9/2}$ orbit is lowered by about 66 keV, while the $f_{5/2}$
orbit is raised by about 66 keV, the effect being opposite for
$^{67}$As. Also the relative energy gap between the proton $g_{9/2}$
and $f_{5/2}$ orbits decreases roughly by 132 keV, providing a large
contribution to the MED. Since the spin-orbit contribution leads to
a reduction of the energy gap between the proton $g_{9/2}$ and
$f_{5/2}$ orbits, excitations from those orbits into the $g_{9/2}$
orbit are enhanced. The opposite effect is predicted to happen in
$^{67}$As for the neutron orbits.

\begin{figure}[t]
\includegraphics[totalheight=7.5cm]{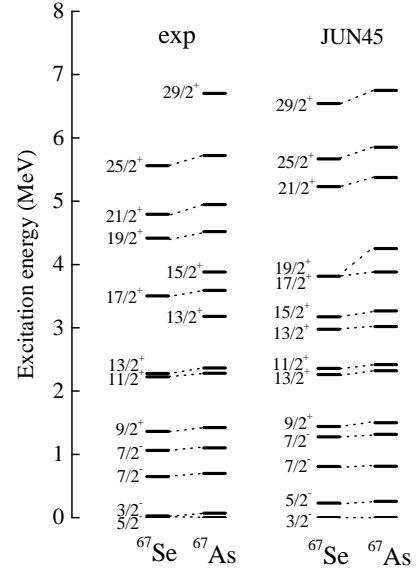}
  \caption{Comparison of calculated energy levels (JUN45) with experimental
  data (exp) for $^{67}$Se and $^{67}$As.}
  \label{fig1}
\end{figure}

With inclusion of $V_{CM}$, $\varepsilon_{ll}$ and $\varepsilon_{ls}$,
shell-model calculations are carried out in the $pf_{5/2}g_{9/2}$
shell for the $A=67$ mirror nuclei.  The isospin nonconserving
term is neglected in the upper half of $fp$ shell region because
the $f_{7/2}$ orbit is almost not active. The calculation uses the
code MSHELL \cite{Mizusaki00} and the effective interaction JUN45.
After solving the eigenvalue problem, contribution of the Coulomb
monopole radial term $V_{Cr}$ is included into the energy $E_{J}$
obtained in the shell model calculation, where the strength parameter
$a_{m}$ was fix to 280 keV so as to fit the MED of the postive-parity
high-spin states, and taken as 0.0 keV for the negative-parity states.

In Fig. \ref{fig1}, the calculated energy levels are shown, and
compared with the experimental data for $^{67}$Se and $^{67}$As.  As
one can see, the calculation with the JUN45 interaction reproduces
well the experimental data.  The energy differences of the analogue
states are in a reasonable agreement with experiment. The structure
of the negative-parity states at low-excitation energies are mainly
dominated by the $fp$ shell configurations. The positive-parity states built
at higher spin strongly involve the $g_{9/2}$ orbit. The structural
difference of such configurations strongly reduces the transition
strengths explaining the isomeric character of the 9/2$^{+}$ levels
\cite{Hasegawa05}. We note however that the calculated level energy
for the 19/2$^{+}$ state lies lower than the experimental value.

In the past few years, experimental data on mirror nuclei above the
doubly magic $^{56}$Ni have become available.  Ekman {\it et al.}
\cite{Ekman04} discussed the MED of the $A\sim 60$ mass region based on the results of
the shell model calculations.  It was reported that for the MED the
contribution of the electromagnetic spin-orbit term
$\varepsilon_{ls}$ is significant, but the monopole Coulomb
$\varepsilon_{ll}$ term is not.
Energy shifts due to $\varepsilon_{ls}$ increase the gap between the $p_{3/2}$,
$f_{5/2}$ and the $g_{9/2}$ orbits for neutrons but reduce it for protons. As a
consequence excitations involving those orbits have important contribution to MED.

\begin{figure}[t]
\includegraphics[totalheight=7.5cm]{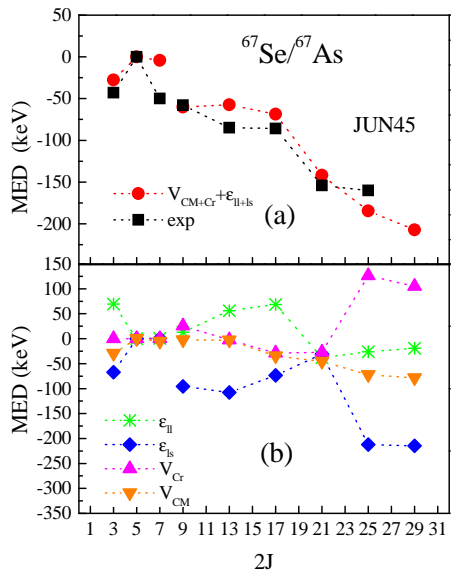}
  \caption{(Color online) The MED for states shown in Fig. 1. Upper graph:
  Comparison of calculated MED with available data. Lower graph: Decomposition
  of theoretical MED into four terms (see text for explanation).}
  \label{fig2}
\end{figure}

In Fig. \ref{fig2} (a), the experimental MED along the
positive-parity excited band with $\Delta J=2$ built on the
$9/2^{+}$ state and the low-lying negative-parity states
($3/2^{-},5/2^{-},7/2^{-})$ are compared with the results of our
JUN45 calculations as a function of spin $2J$.  The agreement is
excellent. In particular, the calculation reproduces correctly the
large negative value in the MED at the high-spin $21/2^{+}$ and
$25/2^{+}$ states.  It is now interesting to examine which terms
contribute to such drastic changes in the MED.  In order to see
this, the four different contributions to MED have been plotted
separately in Fig. \ref{fig2} (b). The Coulomb multipole term
$V_{CM}$ reflects the alignment effects at high spin and follows the
negative trend of the MED. It is in fact well known that spin
alignments affect the MED, which is a behavior first suggested by
Sheikh {\it et al.} \cite{Sheikh90} based on results from the
deformed cranked shell model. In the mirror pair $^{49}$Mn and
$^{49}$Cr the alignment process has been examined based on the
shell-model calculations by counting the number of proton pairs in
the shell $j$ coupled to the maximum spin $J=2j-1$ \cite{Bentley00}.
It was shown that this number is closely correlated with the MED.
For the present case of $^{67}$Se, two protons and one neutron jump
up from the $fp$-shell to $g_{9/2}$ at spin of $25/2^{+}$ and
$29/2^{+}$. The spin alignment of the two protons in the $g_{9/2}$
orbit increases the spatial separation between them, leading to a
smaller Coulomb energy.  Thus, the alignment effect for protons
reduces the excitation energy in $^{67}$Se while the same does not
happen in the analogue states in $^{67}$As.  However, as can be seen
in Fig. \ref{fig2} (b), the $V_{CM}$ term alone underestimates the
MED by a factor of three. As already noticed the contribution of the
$\varepsilon_{ll}$ term is only marginal. $V_{Cr}$ gives the largest
positive contribution in particular for the $25/2^{+}$ and
$29/2^{+}$ states due to the increased occupation of the $g_{9/2}$
orbit. On the other hand, the $\varepsilon_{ls}$ contribution to the
MED is strongly negative for the $25/2^{+}$ and $29/2^{+}$ spin
values. When the $V_{Cr}$, $V_{CM}$, $\varepsilon_{ls}$ and
$\varepsilon_{ll}$ terms are all included, the theoretical MED
reproduce well the experimental data. We note, however, that
the strength of $V_{Cr}$ was fitted to data and not determined in an
independent way. The $\varepsilon_{ls}$ and $V_{Cr}$ terms
contribute to the MED from the opposite directions, causing a large
cancelation at the highest spins.  For the $21/2^{+}$ state, all
terms give almost the same contribution of about 40 keV, providing
in total large MED. Below $21/2$, the $\varepsilon_{ls}$ term
competes with the $\varepsilon_{ll}$ term, while the $V_{Cr}$ and
$V_{CM}$ values are small. In our conclusion therefore the observed
MED behaviour in the $^{67}$Se and $^{67}$As pair is characterized
by a strong competition among the different terms, dominated at high
spin by the interference of the spin orbit and radial contributions.

A question that we now address concerns the importance of the
isospin nonconserving term of the NN interaction. The good results
shown in Fig. \ref{fig2} have been obtained through the inclusion of
the $V_{Cr}$ term whose strength is however fitted to the
experimental data. As seen in Fig. \ref{fig2}, calculations without
$V_{Cr}$ cannot reproduce the data in the high-spin region. In the
$f_{7/2}$ shell nuclei the isospin nonconserving NN term is
important mainly at the low spin region \cite{Bentley07}. If one
speculates that a similar behavior occurs also in the $g_{9/2}$
shell, this would imply a limited contribution of the isospin
nonconserving term to the high spin region in the current
discussion. The calculation presented in this work indeed shows that
we can obtain a good agreement with the experimental data for the
MED without including an explicit isospin breaking NN term. All
these seem to suggest that the isospin nonconserving NN term is not
important. However, since we have normalised a part of the
interaction by fitting to the experimental data, we cannot make a
strong conclusion about the role of the isospin nonconserving part
that in principle contributes to the MED.

\begin{figure}[t]
\includegraphics[totalheight=7.5cm]{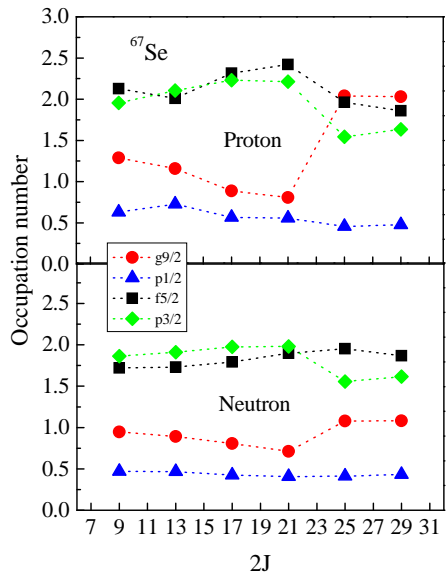}
  \caption{(Color online) Calculated occupation numbers for proton orbits (upper graph)
  and neutron orbits (lower graph) in $^{67}$Se.}
  \label{fig3}
\end{figure}

To support the above picture, Fig. \ref{fig3} shows the calculated
occupancies of the excited band with $\Delta J=2$ built on the
$9/2^{+}$ state in $^{67}$Se.  The upper and lower graphs are for
protons and neutrons, respectively.  From the upper graph, one can
see that for the $9/2^{+}$ state, protons occupy mainly the
$fp$-shell and partially the $g_{9/2}$ orbit.  The occupations
change gradually such that the $fp$ occupancies increase but the
$g_{9/2}$ one decreases as a function of $2J$.  However, it is
notable that the proton $g_{9/2}$ occupation increase suddenly at
spin $25/2$, and the proton $p_{3/2}$ and $f_{5/2}$ occupations
decrease at the same spin.  This means that two protons and one
neutron jump up from
the $fp$-shell to the $g_{9/2}$ orbit at spin $25/2$.  This drastic
change of occupations is in clear contrast to that of the
$f_{7/2}$-shell nuclei, where the occupations of $p_{3/2}$ and
$f_{7/2}$ orbits change gradually with increasing spin \cite{Bentley07}.
The lower graph indicates a similar pattern for
neutrons, but the variation is not so large.  As already mentioned
above, the change in occupancy of the $p_{3/2}$ orbit affects
strongly the MED through the Coulomb monopole radial term $V_{Cr}$.
Since the $p_{3/2}$ orbit has larger radius than the $g_{9/2}$
orbit, when at high spin nucleons are filling the $g_{9/2}$ shell the Coulomb
monopole contribution is larger than that at low
spins.

\begin{figure}[t]
\includegraphics[totalheight=7.5cm]{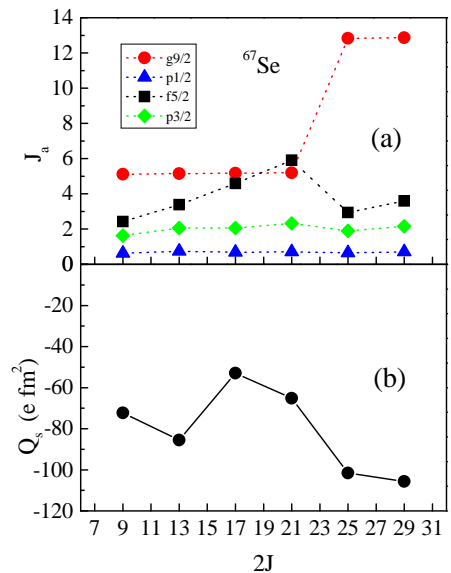}
  \caption{(Color online) Calculated spin distribution in each orbit
  (upper graph) and spectroscopic quadrupole moment (lower graph) in $^{67}$Se.
  The spin distribution is the spin value $J_{a}$ on a single-particle
  orbit $a$ for the excited states (see text for explanation).}
  \label{fig4}
\end{figure}

We finally show the calculated spin alignment and spectroscopic
quadrupole moment in $^{67}$Se.  In Fig. \ref{fig4} (a), the spin
distribution of the expectation value $J_{a}=\sqrt{\langle
\vec{j_{a}} \rangle^{2}}$ is plotted as a function of spin $2J$,
where $\vec{j_{a}}$ is angular momentum operator for each orbit $a$.
As the neutron orbits are blocked for this odd-neutron nucleus, the
first alignment will be that of a pair of $g_{9/2}$ protons which
brings additional 8 units of angular momentum. It is clearly visible
that the proton pair and one neutron alignment in $^{67}$Se
occur at spin $25/2$. The $29/2^{+}$ state also shows a large
aligned spin value. This alignment is interpreted as
five-quasiparticle configuration involving two protons and three
neutrons.  Figure \ref{fig4} (b) shows the calculated spectroscopic
quadrupole moment $Q_{s}$ (in $e$fm$^{2}$) for the excited states
built on the $9/2^{+}$ level in $^{67}$Se. The $Q_{s}$ absolute
value has sudden increase at spin $25/2$ corresponding to the sudden
increase in spin alignment (see the upper graph).  This suggests
that the quadrupole moment is closely related to the spin alignment
of the $g_{9/2}$ proton pair, which correlates with the
multipole term $V_{CM}$ of the MED. Therefore, change in deformation
seems to affect the MED, but its influence is not large. It should
be noted that the present $V_{Cr}$ calculation and the discussion on
occupation of the single-particle orbits are carried out in a
spherical basis, and therefore, the deformation effects (such as
changes of single-particle levels by the shape-driving effect) are
not explicitly seen. To study the deformation effects in the MED, a
shell model based on deformed single-particle states
\cite{Sun1,Sun2} would have to be employed.

In conclusion, we investigated the MED between high-spin states in
the mirror pair $^{67}$Se and $^{67}$As using large-scale shell
model calculations.  The calculations reproduce well the
experimental level schemes, and confirm the suitableness of the
JUN45 effective interaction for this mass region.  The need for
inclusion of the $g_{9/2}$ orbit in the description for the MED in
the upper $fp$ shell nuclei was demonstrated.  In this mass region,
the electromagnetic spin-orbit interaction and the Coulomb monopole
radial term are responsible for producing the large MED at high-spin
states, while the contribution from the Coulomb multipole term is
small.  The occupations of the relevant orbits and the spin
alignment in the $g_{9/2}$ orbit affect the variation of the MED
along the band built on the $9/2^{+}$ state. We obtained a good
agreement with the experimental data for the MED without involving
the isospin nonconserving part. However, it cannot be concluded that
the isospin nonconserving NN term is not important. This remains an
open question. 

Research at SJTU was supported by the National Natural Science
Foundation of China under contract No. 10875077 and by the Chinese
Major State Basic Research Development Program through grant
2007CB815005.  Y.S. thanks a partial support from the Chinese
Academy of Sciences. M.H. thanks support from SJTU-INS Research
Project for Visiting Scholars.




\begin{thebibliography} {99}

\bibitem{Bentley07} M. A. Bentley and S. M. Lenzi, Prog. Part. Nucl.
Phys. {\bf 59}, 497 (2007).

\bibitem{Lenzi06} S. M. Lenzi, J. Phys: Conf. Ser. {\bf 49}, 85 (2006).

\bibitem{Jenkins05} D. G. Jenkins {\it et al.}, Phys. Rev. C {\bf 72}, 031303 (2005).

\bibitem{Nara07} B. S. Nara Singh {\it et al.}, Phys. Rev. C {\bf 75}, 061301 (2007).

\bibitem{Bentley00} M. A. Bentley {\it et al.}, Phys. Rev. C {\bf 62}, 051303 (2000).

\bibitem{Duflo02} J. Duflo and A. P. Zuker, Phys. Rev. C {\bf 66}, 051304(R) (2002).

\bibitem{Lenzi01} S. M. Lenzi {\it et al.}, Phys. Rev. Lett. {\bf 87}, 122501 (2001).

\bibitem{Zuker01} A. P. Zuker {\it et al.}, nucl-th/0104048 (2001).

\bibitem{Garrett07} P. E. Garrett {\it et al.}, Phys. Rev. C {\bf 75}, 014307 (2007).

\bibitem{Garrett01} P. E. Garrett {\it et al.}, Phys. Rev. Lett. {\bf 87}, 132502 (2001).

\bibitem{Gadea06} A. Gadea {\it et al.}, Phys. Rev. Lett.  {\bf 97}, 152501 (2006).

\bibitem{Vedora07} F. Della Vedova {\it et al.}, Phys. Rev. C {\bf 75}, 034317 (2007).

\bibitem{Ekman04} J. Ekman {\it et al.}, Phys. Rev. Lett. {\bf 92}, 132502 (2004).

\bibitem{Bentley06} M. A. Bentley {\it et al.}, Phys. Rev. Lett.  {\it 97}, 132501 (2006).

\bibitem{Brown09} J. R. Brown {\it et al.}, Phys. Rev. C {\bf 80}, 011306 (2009).

\bibitem{Rudolph08} D. Rudolph {\it et al}., Phys. Rev. C {\bf 78}, 021301 (2008).

\bibitem{Andersson05} L. L. Andersson {\it et al.}, Phys. Rev. C {\bf 71}, 011303 (2005).

\bibitem{Warner06} D. D. Warner {\it et al.}, Nat. Phys. {\bf 2}, 311 (2006).

\bibitem{Hasegawa05} M. Hasegawa, Y. Sun, K. Kaneko, and T. Mizusaki, Phys. Lett.
{\bf B617}, 150 (2005).

\bibitem{Honma09} M. Honma, T. Otsuka, T. Mizusaki, and M. Hjorth-Jensen,
 Phys. Rev. C {\bf 80}, 064323 (2009).

\bibitem{Orlandi09} R. Orlandi {\it et al.}, Phys. Rev. Lett. {\bf 103}, 052501 (2009).

\bibitem{Stefanova03} E. A. Stefanova {\it et al.}, Phys. Rev. C {\bf 67}, 054319 (2003).

\bibitem{Jenkins01} D. G. Jenkins {\it et al.}, Phys. Rev. C {\bf 64}, 064311 (2001).

\bibitem{Zuker02} A. P. Zuker, S. M. Lenzi, G. Martinez-Pinedo, and A. Poves,
Phys. Rev. Lett. {\bf 89}, 142502 (2002).

\bibitem{Grzywacz01} R. Grzywacz {\it et al.}, Nucl. Phys. A {\bf 682}, 41c (2001).

\bibitem{Angelis07} G. de Angelis, Prog. Part. Nucl. Phys. {\bf 59}, 409 (2007).



\bibitem{Caurier94} E. Caurier, A. P. Zuker, A. Poves, G. Martinez-Pinedo, Phys.
Rev. C {\bf 50}, 225 (1994).

\bibitem{Nolen69} J. A. Nolen and J. P. Schiffer, Annu. Rev. Nucl. Sci. {\bf 19},
471 (1969).

\bibitem{Mizusaki00} T. Mizusaki, RIKEN Accel. Prog. Rep. {\bf 33}, 14 (2000).

\bibitem{Sheikh90} J. A. Sheikh, P. Van Isacker, D. D. Warner, and J. A. Cameron,
Phys. Lett. {\bf B252}, 314 (1990).

\bibitem{Sun1} Y. Sun, Y.-C. Yang, H.-L. Liu, K. Kaneko, M. Hasegawa, and T. Mizusaki,
Phys. Rev. C {\bf 80}, 054306 (2009).

\bibitem{Sun2} Y.-C. Yang, Y. Sun, K. Kaneko, and M. Hasegawa,
Phys. Rev. C {\bf 82}, 031304(R) (2010).

\end{thebibliography}
\end{document}